\documentclass[twocolumn,prb,showpacs,superscriptaddress,amssymb]{revtex4}

\usepackage{epsfig}
\usepackage{graphicx}
\newcommand{\eg}{\epsilon _\gamma}
\newcommand{\eo}{\epsilon _0}

\begin{document}
\title{Charge dynamics effects in conductance through a large semi-open quantum dot}

\author{Piotr Stefa\'nski}
\email{piotrs@ifmpan.poznan.pl}
 \affiliation{Institute of Molecular Physics, Polish Academy
of Sciences, ul.~M.~Smoluchowskiego~17, 60-179~Pozna\'{n}, Poland}
\author{Arturo Tagliacozzo}
\affiliation{Coherentia-INFM, Unit\'a di Napoli and Dipartimento
di Scienze Fisiche, Universit\'a di Napoli "Federico II", Monte S.
Angelo-via Cintia, I-80126 Napoli, Italy}
\author{Bogdan R. Bu{\l}ka}
\affiliation{Institute of Molecular Physics, Polish Academy of
Sciences, ul.~M.~Smoluchowskiego~17, 60-179~Pozna\'{n}, Poland}
\date{Received \hspace{5mm} }
\date{\today}

\begin{abstract}
Fano  lineshapes  in resonant  transmission  in  a quantum  dot
 imply  interference  between
 localized and   extended  states.
 The influence of the charge
accumulated at the  localized levels, which
 screens the external gate voltage
acting  on  the  conduction  channel is investigated.
 The modified Fano $q$ parameter and  the  resonant  conduction
 is derived starting from a microscopic
Hamiltonian. The latest experiments on "charge sensing"  and
  ``Coulomb  modified  Fano  sensing ``
compare  well with  the  results  of  the present model.

\end{abstract}
 \maketitle
\section{Introduction}
Conductance  peaks  and  Coulomb  Blockade (CB) valleys  mark
  the charge  transport    across   a small
quantum dot (QD), weakly  coupled  to  the  contacts, when   the
 gate  voltage  $V_g $  is  varied.
However, in  a large  QD the multilevel  structure cannot be
ignored. The  dot  may  display   a ``core''  level structure
including
  levels $\eg $  which  are   localized  at  different  places  of the
dot  area. Hence,
  they  may  not
 hybridize   among  themselves, nor  directly  with  the  contact leads,
 but  they may
be   weakly  coupled to  one  or
 more conduction channels.  The  latter could be  of  ``edge''  type
and extend  from  the left (L)  to  the right (R) contact.  The
existence of such   an edge and core level structure in large
QDs has been proven experimentally by Single Electron
Capacitance Spectroscopy \cite{ashoori}.
In   linear  transport measurements,
 the coupling of localized  and  delocalized
 states, called "bouncing states" in \cite{lindermann},
 has  been  invoked  to  justify   the asymmetry of the CB peaks.

 On  the  other  hand,  Fano  resonances
are   ubiquitous   in  the conductance   of nanodevices.
They  arise  when  localized  levels  are  embedded  in
  a continuum  of single  particle
 extended   states  which  contribute  to  transport \cite{fano}.
 Linear  conductance on  a semi-open  QD displays many
transmission   peaks  with the  characteristic  Fano  shape,
 each  time  a core level  crosses  the chemical  potential
 $\mu $ \cite{fuhner}.

 In  a previous  paper \cite{psatbb}
we  have  shown  that a  Fano-like   pattern   also  appears when
  the continuum
  of   conduction  states  at    $\mu $  originates  from  electron
 correlations,  as  is  the  case  of   a broad Kondo
  resonance \cite{gordon},
 due to   a  strongly  hybridized  level   $\eo$,
lying deeply  below  $\mu $. A  bunch  of localized  core levels
$\eg $,  weakly  coupled  to the $\eo$ level,  imprints
 the  broad  Kondo  peak    with    Fano  lineshapes.
 The Kondo  conductance
  tends  to  freeze  the charge sitting  on  the  $\eo$ level,
  in  competition
 with  the  Fano  mechanism which causes particle number oscillations.
 Kondo correlations  stick  the  charge  occupancy  of  the $\eo$
level to  the  value one,  and   the  Fano  factor  $q$  close  to
zero. It  follows that   the transmission  shows  a  dip  in  form
of an  antiresonance.

In  opening  the  QD  further,  by  increasing  the  coupling to
  the  contacts, $t_r$ (r= L,R ),  the  CB structure  is  eventually
  washed  out.
We  assume  that  a  single edge  channel  of  energy  $\eo$  is
centered  at  $\mu$,
 providing   a broad   single  particle  resonance  peak in the transmission.
  The electron  flow  across the
dot  is  essentially   uncorrelated.
  Nevertheless,   the dynamics of the  charge   in  the  core
 levels  does  influence the  resonant  tunnelling, as we  show  in
  Section II  of  the  present  work.

  The $\eg $ levels  hybridize  with  the  contacts
 only  indirectly,  by weakly  coupling to  the  $\eo$  resonance,
   with  a hopping   matrix  element $t_\gamma$.
We  also include   some     on-site  Coulomb  repulsion  $U$  at
the  core
 sites,    which  produces   a  lower  Hubbard
level   $\eg$ (with  single  occupancy)   and  an  upper  Hubbard  level
(with double  occupancy)   $\eg + U $.

  By  including   the
charge  piled up in  the $\eg$ states  within  the  Hartree-Fock
(HF) approximation, we  find  that  the  main  features  of   the
Fano landscape   are not lost. There  are  two  main
  alterations, however :
 $ a)$  the  charge  screens  the gate  voltage $V_g $ by acting
  on  the  $\eo$  level
 with  a capacitive interaction  and it shifts  the  resonances;
$ b) $   the  presence of  the  charge   modifies  the   Fano  $q$ factor
 and  induces  a  marked   asymmetry  in  the  peaks.

In  Section III  we  elaborate  on  modeling  the   experimental
  setup  that  has  been investigated  recently by Johnson \textit{et al.}
 \cite{marcus}. The
  conductance   measured  through a quantum point contact (QPC)
 "senses" the   addition of   charge  to     a  QD,
 which is  capacitively coupled,   parallel to  the  QPC.
 We  show  that our  model
  applies  with  small  modifications.  In this case the
edge channel is not internal to the dot but it describes the QPC
conductance, while the $\eg$ levels localized
 within the dot are  fed with  charge  by  an  external  reservoir.
We  neglect  $U$ which  implies  only  minor  modification  when  the
static  screening  approximation  holds,  but  we include a small
hybridization
$t_\gamma $  of  the  core states to  the current  carrying  QPC state.
This  allows  us  to  monitor  the  full  range  from  bare  capacitive
  coupling  or ``charge  sensing `` (CS), to  Fano  resonant  tunneling.

An increasing gate voltage applied to the dot shifts the  discrete
  structure  of  energy  levels across  $\mu$
 and the core levels are gradually filled by electrons.
The  charge   accumulated in the  core levels
  capacitatively  modifies
the gate voltage acting on the conducting channel.
In the  ``charge   sensing ``  case,
   sudden   jumps  appear
in the  conductance,  monitoring  the  discreteness of  the
charge  accumulation  process.
 Additionally, when $t_{\gamma}\neq$ 0, a set of Fano resonances emerges in the
conductance across the dot as a result of quantum interference. We
show that the shape of Fano resonances is markedly affected by the
charge  trapping within the core levels  as found in  the
  experiment\cite{marcus}.

\section{model Hamiltonian of the Quantum Dot and calculations}
The Hamiltonian of the system under consideration is composed of
parts describing the edge state of the dot, $H_{edge}$, the core
QD levels, $H_{core}$ and the  electrodes $H_{el}$:
\begin{widetext}
\begin{eqnarray}
\label{for1}
H= H_{edge}+H_{core}+H_{el}, \nonumber\\
H_{edge}=\sum_{\sigma}\left (
\epsilon_{0} -V_g + (U'/2) \sum _{\gamma \sigma ' }
n_{\gamma\sigma '} \right ) a_{\sigma}^{\dagger}a_{\sigma} +\sum_{k\sigma
\alpha}t_{\alpha}(V_g)\lbrack c_{k\alpha \sigma}^{\dagger}
a_{\sigma}+h.c \rbrack,\nonumber\\
H_{core} =
\sum_{\gamma\sigma}\left (\epsilon_{\gamma} - V_g \right )
d_{\gamma\sigma}^{\dagger}  d_{\gamma\sigma}+\sum_{\gamma}
Un_{\gamma\uparrow}n_{\gamma\downarrow}
 + \sum_{\gamma\sigma}t_{\gamma}\lbrack
 d_{\gamma\sigma}^{\dagger}a_{\sigma}+h.c.
 \rbrack , \nonumber\\
H_{el}= \sum_{k\sigma,r=L,R}\epsilon_{kr}c_{kr
\sigma}^{\dagger}c_{kr\sigma}.
\end{eqnarray}
\end{widetext}
The edge QD state  of  energy   $\epsilon_0$  is hybridized with
electrodes. The matrix element $t_{r}$, ($r=$L, R) has a given
dependence on the gate voltage $V_g$,  very  much like what
happens  in  a  QPC,
 which opens up when the gate
voltage increases \cite{cronenwett}.  The  capacitive  coupling  is
described  by  a cross-interaction $U'$ between  the  edge
 and  the  core  states  ($ n_{\gamma \sigma } =
 d_{\gamma \sigma}^{\dagger}d_{\gamma \sigma} $
 is  the  occupation  number  on  the core
 state $\mid\gamma \sigma\rangle$ ).

The  retarded  Green  function  for  the  edge   state  is
obtained  by the  Equation  of  Motion  Method  (EOM) with  an
additional HF decoupling  of  the  capacitive interaction:
\begin{widetext}
\begin{equation}
\label{g0}
 G_{0,\sigma}
\left (\omega , V_g ; \langle n_{\gamma\bar{\sigma}}\rangle \right
)=\left \{ \langle 0 |\lbrack \: \omega -  H_{edge} ( V_g ;
\langle n_{\gamma\bar{\sigma}}\rangle ) \rbrack ^{-1} | 0 \rangle
-\sum_{\gamma}\frac{t_{\gamma}^2}
{\omega-\epsilon_{\gamma,\sigma}+ V_g} \right \} ^{-1} \:\: .
\end{equation}
\end{widetext}
The  first  term  on  the  r.h.s. is  an  average  on  the  ground  state
of  the  system $|0 \rangle $ (zero  temperature  is  assumed).
It  includes  hybridization  with the
  contact  leads. The  occupancy  of  the core  levels  is
  selfconsistently  determined from
 $\langle  n_{\sigma} \rangle
=(-1/\pi)\int \Im m G_{\gamma,\sigma}(\omega)d\omega $, where
 the Green
function of a $\gamma$-level with spin $\sigma$ is given in the
Hubbard  approximation \cite{hewsonart}:
\begin{equation}
\label{gamma_hub} G_{\gamma,\sigma}(\omega)=\left [
\frac{(\omega-\tilde{\epsilon}_{\gamma,\sigma})(\omega-
\tilde{\epsilon}_{\gamma,\sigma}-U)}
{\omega-\tilde{\epsilon}_{\gamma,\sigma}-U(1-\langle
n_{\gamma\bar{\sigma}}\rangle)} + i\Gamma_{\gamma,\sigma}\right ]
^{-1}.
\end{equation}
Thus, $G_{\gamma,\sigma}$ has finite width
\begin{equation}
\Gamma_{\gamma,\sigma} = -\Im m \sum_{\gamma}   \: t_{\gamma} \;
\langle 0 |\lbrack \: \omega -  H_{edge} ( V_g ;  \langle
n_{\gamma\bar{\sigma}}\rangle ) \rbrack ^{-1} | 0 \rangle  \;
t_{\gamma} \label{Ggamma}
\end{equation}
due to indirect  interaction with electrodes.
The bare $  \eg  -V_g $ is also
additionally shifted  $ \to  \tilde\epsilon_{\gamma}$
by the real part of the r.h.s. in  Eq. \ref{Ggamma}.
 Charge
accumulation in the core states themselves shifts  the
position of these levels with respect to the chemical potential
even  further,  but  we neglect this capacitative
modification of  their energy  location  in  the  calculation,
   since  it  is  inessential,  as  core  levels  do not
participate directly in the transport across the dot.

The linear conductance
 (i.e. in zero limit of the drain-source voltage) in  units  of  the  quantum  conductance  $2e^2 /\hbar $,
is \cite{meir}:
 \begin{eqnarray}
 \label{cond}
 \mathcal{G}=\sum_{\sigma}\int_{-\infty}^{\infty}\Gamma_{\sigma}(\epsilon)\left(-\frac{\partial f(\epsilon)}{\partial
 \epsilon}\right) \; \rho_{0,\sigma}
\left (\omega , V_g ; \langle n_{\gamma\bar{\sigma}}\rangle \right
)d\epsilon,
 \end{eqnarray}
where $f(\epsilon)$  is the Fermi distribution function,
$\Gamma_{\sigma}(\epsilon)=\Gamma_{L\sigma}(\epsilon)\Gamma_{R\sigma}
(\epsilon)/[\Gamma_{L\sigma}(\epsilon)
+\Gamma_{R\sigma}(\epsilon)]$  is the  hybridization  of  the edge
state with  the  contact leads ($\Gamma_L$= $\Gamma_R$ is further
assumed) and the spectral density $\rho_{0,\sigma}=-(1/\pi)\Im  m
G_{0,\sigma}$.

It  is  instructive   to  introduce  Fano parameter  $q$ defined
as
 \cite{stone,stefanski}:
\begin{equation}
\label{fanoq} q_{\sigma}(V_g)=-\frac{\Re e
G_{0,\sigma}\left (\omega=0,V_g ; \langle n_{\gamma\bar{\sigma}}\rangle
\right )}{\Im m G_{0,\sigma}\left ( \omega=0, V_g ; \langle
n_{\gamma\bar{\sigma}}\rangle \right )},
\end{equation}
with gate voltage  screened by the trapped charge:
\begin{equation}
\label{gateint}
 V_g^{SC}=V_g-
 (  U' /2) \sum_{\gamma \sigma}\langle n_{\gamma\sigma}(V_g)\rangle,
\end{equation}

\begin{figure} [ht]
 \epsfxsize=0.4\textwidth \epsfbox{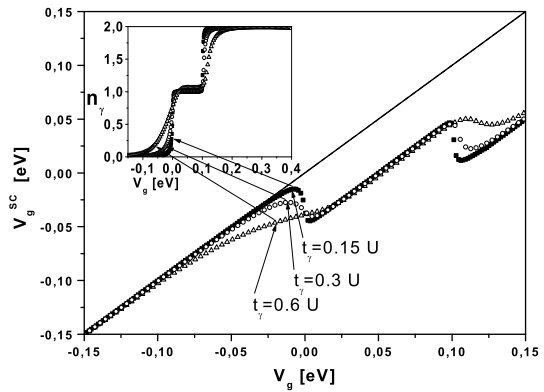}
\caption{\label{gate}Gate voltage  screened  by the filling of the
$\gamma$ level when  the latter  is  moved  below $\mu =0
$ (  $\alpha =1 $).
 The inset shows the  selfconsistently calculated total occupancy
 $n_{\gamma}$.
  The straight line is  the  bare $V_g$ and is  drawn
for comparison.}
\end{figure}

Let  us  consider  one  single  core  level  for  the  time being.
The  $V_g$, $V_g^{SC}$ are  the bare  and screened voltage acting
on the  edge level,  but  they   move the core level  as well. In
crossing    $\mu $, the core  level is
 gradually  filled with electrons. The occupancy of the $\gamma$  level
is  shown  in the inset of  Fig.(\ref{gate}) for  various
$t_\gamma $'s. In  turn, while  the  occupancy  of  the  core
level   changes,
 a  step appears  in  the  screened  voltage, resulting  from   the  capacitive
coupling between  the  core  and  the  edge state.

$V_g^{SC} $  is  plotted   in   Fig. (\ref{gate}) {\sl vs }  the
bare $V_g$. Two  such  kinks appear,  corresponding  to  the
single and  double occupancy  of  the  $\gamma$ level. Their
energy separation is  of the  order of the Coulomb repulsion $U $
(taken as $U= 0.1 \; eV $ in  the  calculation).
  The   dependence of  the  hybridization  parameter
on  the voltage  is implemented   by multiplying the matrix
element
 $t_{r}^{max} = 2 U $ by
the function $\theta (V_g)=\lbrace exp[(s
-V_g)/\delta]+1\rbrace^{-1}$, where $\delta$ (= 0.5) controls the
sharpness of the kink, and $s$ (= -0.2 in calculations) its
position. The kink is smoothed with   increasing of  the
hybridization $t_\gamma $ between $\epsilon_0$ and
$\epsilon_{\gamma}$ because the electron has an increasing
preference to tunnel resonantly through $\epsilon_0$ instead of
dwelling on the $\epsilon_{\gamma}$ level\cite{psatbb} when
$\epsilon_0$ crosses the  chemical potential.
  Here  $\alpha  =  U'/U $ is the charge sensitivity parameter.
 It
  can increase to  unity when the level broadening in QD becomes larger
than the level  spacing and the cross-interaction $U'$ approaches
 on-site Coulomb  repulsion $U$.

\begin{figure} [ht]
\epsfxsize=0.4\textwidth \epsfbox{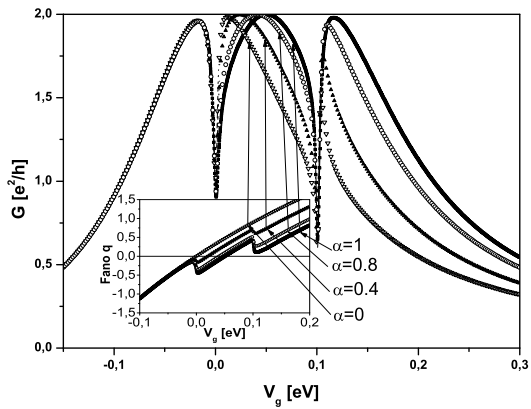}
\caption{\label{fanohub}Conductance {\sl vs} $V_g $ for
 $t_{\gamma}=0.15 \; U $ at various charge  sensitivities  $\alpha$ of the
filling of the $\gamma$-level. In  the  inset the  Fano $q$, Eq.
(\ref{fanoq}), is shown.}
\end{figure}

In Fig. (\ref{fanohub}) we  plot  the  corresponding  conductance,
 for various $\alpha$.
The filling of the $\epsilon_{\gamma}$ level  has a  marked effect
on the shape of the Fano resonances because it modifies  the Fano
$q$ factor (see inset). The non  zero  value of $q$  and its  sign
are  responsible  for the  asymmetry of the Fano dips within  the
broad  conductance  peak  centered  at $\epsilon_0  \sim  0 $.

\section{charge and Fano ``sensing" in  the  QPC conductance}

Our model can also be applied straightforwardly  to
describe  recent transport measurements through a QPC
 ``sensing" the  charge  piled  up  in  a
 QD  placed  at its   side  and  capacitively  coupled  to  it
\cite{marcus}.  We  show  that  it  is  possible  to  continuously  move
 from  a ``capacitive  sensing''  to  a ``Fano  sensing''  by  increasing
the  hybridization  of  the  single  particle  levels  $\eg $  of the  dot
with  the conduction channel  $\eo$  of  the  QPC.

\begin{figure} [th]
\epsfxsize=0.4\textwidth \epsfbox{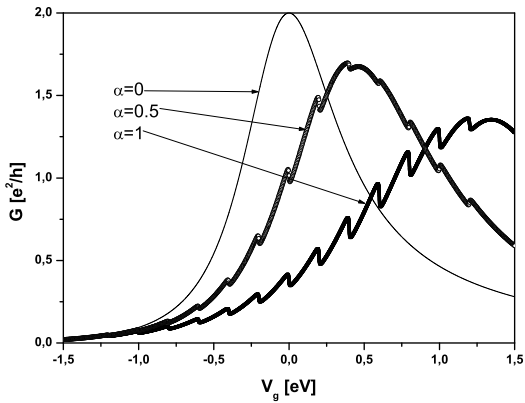}
\caption{\label{QPCcharge}Conductance through the QPC with "charge
sensing" only ($t_{\gamma}=$0) for various $\alpha$ parameter and
$t_{B}/\Delta$=0.15. The curve for $\alpha$=0 without charge
sensing is also shown for comparison.}
\end{figure}

\begin{figure} [t]
\epsfxsize=0.4\textwidth \epsfbox{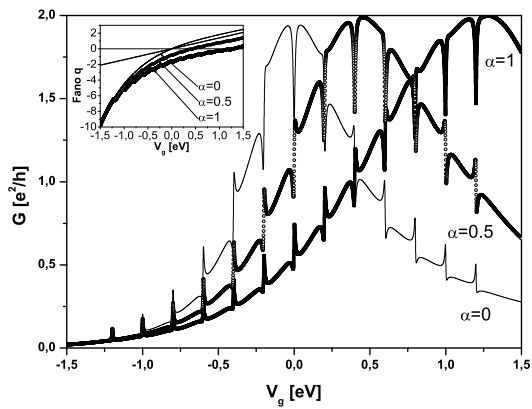}
\caption{\label{QPCfano} Conductance through the QPC with "charge
sensing" and finite hybridization with the dot ($t_{\gamma}$=
0.15$\Delta$) for various $\alpha$ parameter. Inset shows
corresponding Fano $q$, the straight line is for $\alpha$=0 and
constant coupling to electrodes $t_{r}=t_{r}^{max}.$}
\end{figure}

To  match  with  the  experimental  setup  we  use  the
Hamiltonian  of Eq. (\ref{for1})  by  reinterpreting  it  as
follows:

$H_{dot} \equiv H_{core}+ \sum_{k,\sigma}\epsilon_{k}^B
b^{\dagger}_{k\sigma} b_{k\sigma}+
\sum_{k\gamma\sigma}t_{B}[b^{\dagger}_{k\sigma}
d_{\gamma\sigma}+h.c]$ and $H_{QPC} \equiv H_{edge}+ H_{el}$. We
have   added an  additional  reservoir $B$, in  contact  with the
dot,
 with  levels $\epsilon_{k}^B $ corresponding  to  single  particle
operators  $ b_{k\sigma}$. A featureless
 and broad density of states   for  the  reservoir is assumed.  The  dot  is  supplied  with  electrons
 from the bath $B$ due to the  hopping  parameter $t_B$. The level  $\epsilon_0$
corresponds now to the QPC, having a  gate voltage dependent
hybridization with the leads. In  Figs. (\ref{QPCcharge},
\ref{QPCfano}) we  report the results  for a  multilevel   dot
with $\gamma$= 13 discrete levels equally  spaced  by  $\Delta
=0.2 \; eV$
 and symmetrically  located  with respect to $\mu =0 $ for $V_g =0 $. In  this
  hypothesis  we can  assume  that  $\Delta $  is  the
   addition  energy of  the dot,  independent  of the  electron  number
$N$. Thus,  we
drop  the  interaction  term $\propto U $ in  $H_{core}$ and the
relevant  energy  scale  for  charge  sensing  is  now  $\Delta $  itself.
 A static  screening in  the  multi-electron system  is  expected  to  be
 quite   effective,  if  the  dot  is  large.
The screened  gate voltage acting on the  QPC, sensing  the  charge  on
 the  dot,  takes  now  the  same  form as Eq. (\ref{gateint}),
 but with $U'$ replaced
by the charging energy of the dot $\Delta$.
 The $\alpha$ parameter now describes
the  sensitivity  of  the capacitive coupling between the QD and
QPC which can be controlled by   appropriate gates \cite{marcus}.
Again,  the particle number on the dot
$\sum_{\gamma,\sigma}n_{\gamma,\sigma}(V_g)$ is
   determined by  means of  the  Friedel sum rule
\cite{hewson}.  It  shows  the  usual  steps  as  in  the  inset
of Fig. (\ref{gate})  when the gate voltage increases and the dot
is filled by electrons.
 For $\alpha$=1 the CS is
equal to the charging energy, each time an electron hops onto the dot.
 Fig. (\ref{QPCcharge}) shows the conductance through the  QPC when
the dot is only capacitively coupled to it ($t_{\gamma}$=0), for
various $\alpha$'s.
 For $\alpha$=0   the dot  has  no  effect  on  the QPC  conductance.
The  transmission  across  the QPC displays  a broad  resonance
centered  at   $\eo  =\; V_g =\; 0$. Its  width  and
characteristic asymmetry is governed by the gate voltage dependent
coupling to the $L,R$ contacts. By  increasing $\alpha$, the
maximum of the resonance is  moved to  higher $V_g$'s, because of
the
 voltage screening. The conductance  peak  acquires a  sawtooth
pattern,  which grows sharper when $\alpha$ increases.  The
pattern follows
 the jumps  of  $V_g^{sc}$, each  time  a new  electron  enters  the  dot.
 The shape of the overall conductance curve is
similar  to the  one  seen in the experiment (see Fig. (2a) of
\cite{marcus}).

In Fig.(\ref{QPCfano}), the conductance is reported  for finite
hopping between QD and QPC ($t_{\gamma}$=0.15 $\Delta$), at
$\alpha =$0.0, 0.5, 1.0. The  case $\alpha =0 $ is  the  pure Fano
case. The  maximum of  the conductance is  still  centered at $V_g
\approx 0$  because  $t_\gamma  $  has  been  chosen  to be rather
small. The zero  of  the  real  part of the  QPC Green's function
$G_{0,\sigma }$
  defines
 the  location  of  the maximum  of  the resonance, and  the  zero  of  the
Fano $q$  as well, according  to  Eq.(\ref{fanoq}).
 The symmetric Fano dip corresponding  to $q=0$
 is  centered on top  of  the  conductance maximum, because a
$\gamma $  level  sits  exactly  at  $V_g =0 $.
 When  $\alpha $ increases,  the  condition  $ \eo -V_g^{SC} \sim 0 $
  now   marks  the location  of the conductance  maximum,  as  well  as
the  place  where $q \sim  0$ (see  the inset of
Fig.(\ref{QPCfano})). Thus, the conductance envelope    moves
again to  larger voltages,  together with  the  Fano-like
structures, and  the  conductance  peak  broadens as in Fig.
(\ref{QPCcharge}). Appearance of sharp peaks for large negative
gate voltages is caused by rapid decrease of $q$ (see inset) due
to the closing of QPC, Eq. (\ref{fanoq}).  Our  case with $\alpha
=1.0 $  compares  well with  Fig.(2b) of \cite{marcus}, displaying
Coulomb modified  Fano  resonances. Both ref. \cite{marcus} and in
ref.\cite{berkovits} propose a model
  conductance based  on  the  grand canonical  average  of  dot
 configurations, which  can  be  fitted  to  the  experimental  data.
Our  calculation  shows  that  the  dynamics  of  the  resonant
tunnelling together  with  some  minor assumptions provide simpler
results,
  in qualitative
agreement  with  the  experimental  data.

To  conclude,  we  have  demonstrated  that  a  Coulomb  modified  Fano
  resonant  tunneling  arises from charge  sensing in  structures like
  a QPC  with  a dot  aside,
 or inside  a large  dot  where  core
levels coexist  with  extended  edge  states  and  both  can  be  described
within  the  same  quantum  mechanical  model  with  minor  modifications.

\begin{acknowledgements} This work was supported by the Polish
State Committee for Scientific Research (KBN) under Grant no.
PBZ/KBN/044/P02/2001 and the Centre of Excellence for Magnetic and
Molecular Materials for Future Electronics within the European
Commission Contract No. G5MA-CT-2002-04049.
\end{acknowledgements}

\end{document}